\documentclass[a4paper,fleqn,usenatbib,letters,useAMS]{mnras}

\usepackage{newtxtext,newtxmath}

\usepackage[T1]{fontenc}
\usepackage{ae,aecompl}

\usepackage{graphicx}	
\usepackage{amsmath}	
\usepackage{amssymb}	
\usepackage{amsfonts}
\usepackage{epstopdf}
\usepackage{comment}

\title[3PCF randoms counts w/o randoms]{Computing 3 point correlation function
randoms counts without the randoms catalogue}

\author[D. W. Pearson \& L. Samushia]{
David W. Pearson,$^{1}$\thanks{E-mail: dpearson@phys.ksu.edu} \&
Lado Samushia$^{1,2}$
\\
$^{1}$Department of Physics, Kansas State University, 116 Cardwell Hall,
Manhattan, KS, 66506, USA\\
$^{2}$National Abastumani Astrophysical Observatory, Ilia State University, 2A
Kazbegi Ave., GE-1060 Tbilisi, Georgia\\
}

\date{Accepted 2019 May 01. Received 2019 April 23; in original form 2019 March 22}

\pubyear{2019}

\begin{document}
\label{firstpage}
\pagerange{\pageref{firstpage}--\pageref{lastpage}}
\maketitle

\begin{abstract} 
As we move towards future galaxy surveys, the three-point statistics will be
increasingly leveraged to enhance the constraining power of the data on
cosmological parameters. An essential part of the three-point function
estimation is performing triplet counts of synthetic data points in random
catalogues. Since triplet counting algorithms scale at best as
$\mathcal{O}(N^2\log N)$ with the number of particles and the random
catalogues are typically at least 50 times denser than the data; this tends to be
by far the most time-consuming part of the measurements. Here we present a
simple method of computing the necessary triplet counts involving uniform random distributions through simple one-dimensional integrals. The method speeds up the computation of the three-point function by orders of magnitude, eliminating the need for random catalogues, with the simultaneous pair and triplet counting of the data points alone being sufficient.
\end{abstract}




\section{Introduction}
Current and past galaxy redshifts surveys have heavily relied upon the analysis
of two-point statistics to constrain cosmological parameters down to
the percent level \citep[see e.g.][]{Anderson2012,Chuang2016,Cuesta2016,GilMarin2016a,
GilMarin2016b,Alam2017}. 
As upcoming galaxy redshift surveys seek to push
constraints on cosmological parameters to the sub-percent level, three-point
statistics -- the three-point correlation function (3PCF) and the bispectrum --
will begin to play a bigger role in analyses. Recent works have shown that the
baryon acoustic oscillation features are detectable in both the 3PCF
\citep{Slepian2015,Slepian2017a,Slepian2017b} and the bispectrum
\citep{Pearson2018,Pearson2019}, hinting at the possibility of increased
constraining power on cosmological parameters via the inclusion of three-point
statistics. The large scale bispectrum has also been used to supplement the
redshift-space distortion measurements from the power spectrum 
\citep{GilMarin2015a,GilMarin2015b,GilMarin2017}. 

Turning to small scales, the Halo Occupation Distribution (HOD) model 
\citep[and references therein]
{Scherrer1991,Jing1998,Peacock2000,Seljak2000,Scoccimarro2001,Berlind2002,Zheng2009} 
is a popular method of linking
the galaxy and dark matter distributions \citep[see e.g.][]{Tinker2005,Guo2014,Guo2016,
Rodriguez-Torres2016}. \cite*{Yuan2018} have shown
that the squeezed 3PCF can help tightly constrain the parameters of the HOD, 
making it very likely that the 3 point statistics 
will become important to the development of mock catalogs.

To measure the 3PCF from data, one must count the triangles of specific 
shapes and sizes from the data  -- e.g. the galaxy catalogue -- as well 
as a set of unclustered random points -- the random catalogue -- a process which 
is naively $\mathcal{O}(N^3)$ in time complexity. The counts from the 
data are then compared to the expected mean numbers from the unclustered 
random points to get an estimate of the 3PCF \citep{Peebles1975,Peebles1980}, 
as the unclustered mean triplet counts are very sensitive to the geometry 
of the survey and its number density variations. The most popular
estimator used for the 3PCF is that of \cite{Szapudi1998} due to its 
superior edge effect corrections \citep{Kayo2004},
\begin{equation} 
\label{eq:estimator}
\begin{aligned} 
\zeta(r_{1},r_{2},r_{3}) =&\;
\dfrac{1}{RRR(r_{1},r_{2},r_{3})}\left[DDD(r_{1},r_{2},r_{3}) -
3DDR(r_{1},r_{2},r_{3})\right.\\ &\;\left.+\; 3DRR(r_{1},r_{2},r_{3}) -
RRR(r_{1},r_{2},r_{3})\right], 
\end{aligned} 
\end{equation}
where the combinations of $D$s and $R$s tell you how many vertices of the 
triangle come from either the data or the random points, respectively. 
The random catalogue usually contains 50 or more times as many objects as 
the data to make the shot-noise in this Monte-Carlo estimation subdominant 
to the variance in the data. Combine this with the $\mathcal{O}(N^3)$ 
complexity and the counts for $DDR$, $DRR$, and $RRR$ can consume a large 
amount of time.

While the computational complexity of the three-point statistics can be somewhat mitigated \citep{Baldauf2015,Slepian2015,Pearson2018}, their calculation still tends to take a significant amount of CPU time. 
Additionally, the studies which stand to benefit from the inclusion of the 
three-point statistics may need to measure them tens of thousands of times, 
making any potential reduction in the computational complexity a welcome 
improvement.

In this letter, we present a simple method of obtaining the counts involving
the random points that does not require actually counting triangles from
random catalogues. Since triplet-counting from random catalogues takes most
of the computational time, this results in a significant reduction in required CPU
hours. The method is only applicable to measurements of 3PCF from uniform
periodic cubes, which means that, unfortunately, it does not apply to
measurements from real survey data. There are, however, many stages in the
cosmological analysis of the 3PCF that can be performed on uniform periodic cubes
and do not need to account for survey geometry. These include the
validation of theoretical templates -- comparing model predictions to
measurements from simulations and quantifying biases -- and the HOD parameter
fitting. For these applications our method allows one to set the
number density of randoms arbitrarily high with no performance degradation.

\section{Method}

\subsection{Pair counts}

We start by reminding the reader why random pair counting is not really
necessary when computing the two-point statistics from a periodic cube 
\citep[see][for an application of analytic two-point randoms counts to the 
angular correlation function]{Roukema1994}. The number of
unclustered pairs separated by a distance $r \pm \Delta r/2$ can be easily
computed analytically. There are on average
$n_{R}V_{\mathrm{box}}$ particles in a uniform periodic cube with volume 
$V_{\mathrm{box}}$ and a number density of
points $n_{R}$. The volume between two concentric spheres of radii $r - \Delta
r/2$ and $r + \Delta r/2$ -- i.e. a spherical shell -- is 
\begin{equation}
\label{eq:shell}
V_\mathrm{shell} = \frac{4\upi}{3}\left[\left(r + \frac{\Delta r}{2}\right)^3 -
\left(r - \frac{\Delta r}{2}\right)^3\right],
\end{equation}
\noindent
and the total number of pairs is
\begin{equation}
\label{eq:pairs}
N_\mathrm{pairs} = n_{R}^{2}V_{\mathrm{box}}V_\mathrm{shell},
\end{equation}
\noindent
since The cube is periodic there are no edge effects. We will now generalize
this idea for triplet counts.

\subsection{RRR counts}
\label{subsec:RRR}
We need to estimate the number of triplets separated by $(r_{1} \pm \Delta r/2,r_
{2} \pm \Delta r/2,r_{3} \pm \Delta r/2)$ in a uniform periodic cube with a
number density of $n_{R}$ and volume $V_{\mathrm{box}}$. We will start by 
locating the number of $p_3$ points given $p_1$ and $p_2$, which is achieved 
by finding the volume of the two overlapping spherical shells whose cross-sections 
are shown in Figure 1, and multiplying by our number density.

\begin{figure}
\includegraphics[width=\linewidth]{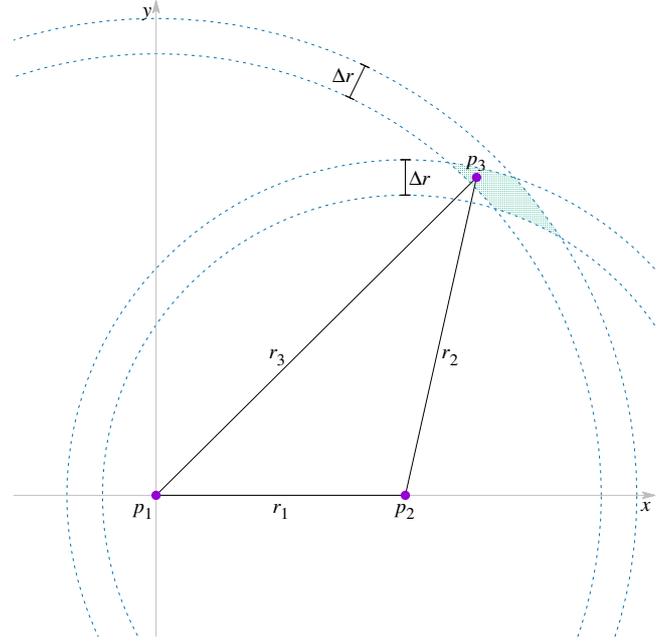}
\caption{Diagram showing the volume of interest for determining the number of
triangles expected given points $p_{1}$ and $p_{2}$. In essence, we want to
predict how many points, $p_{3}$, there will be, which is simply the volume
of the overlap region of the two spherical shells with central radii $r_{2}$
and $r_{3}$, multiplied by the number density.}
\label{fig:overlap}
\end{figure}

We will make use of the equation for the volume of  overlap of two spheres
with radii $r$ and $R$ separated by a distance $d$ \citep{Weisstein},
\begin{equation}
\label{eq:sphereSphere}
    \mathcal{V}(d, r, R) = \dfrac{\upi(R + r - d)^{2}(d^{2} + 2dr - 3r^{2} + 2dR + 6rR - 3R^{2})}
                                     {12d}.
\end{equation}
This equation is only valid if the spheres touch at one or more points, i.e. $R - r \leq d \leq R + r$. 
Because of this, we need to
explicitly define the volume outside of those bounds,
\begin{equation}
\label{eq:sphereSphereGen}
    V(d, r, R) = \left\lbrace\begin{array}{lr}
                        0 & d > R + r \\
                        \mathcal{V}(d,r,R) & R - r \leq d \leq R + r \\
                        \dfrac{4}{3}\upi r^{3} & d < R - r \\
                        \end{array}.\right.
\end{equation}

\begin{figure}
\includegraphics[width=0.49\linewidth]{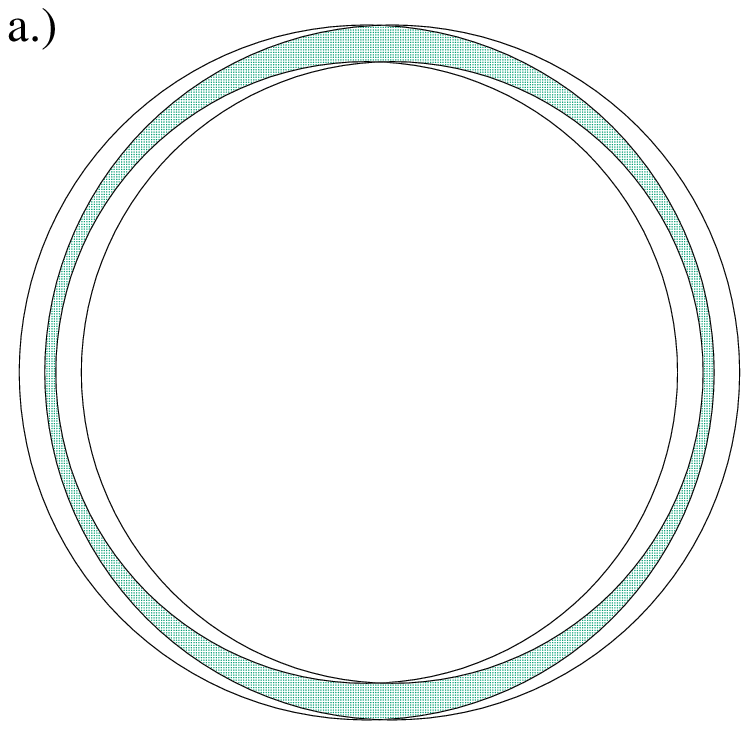}\hfill
\includegraphics[width=0.49\linewidth]{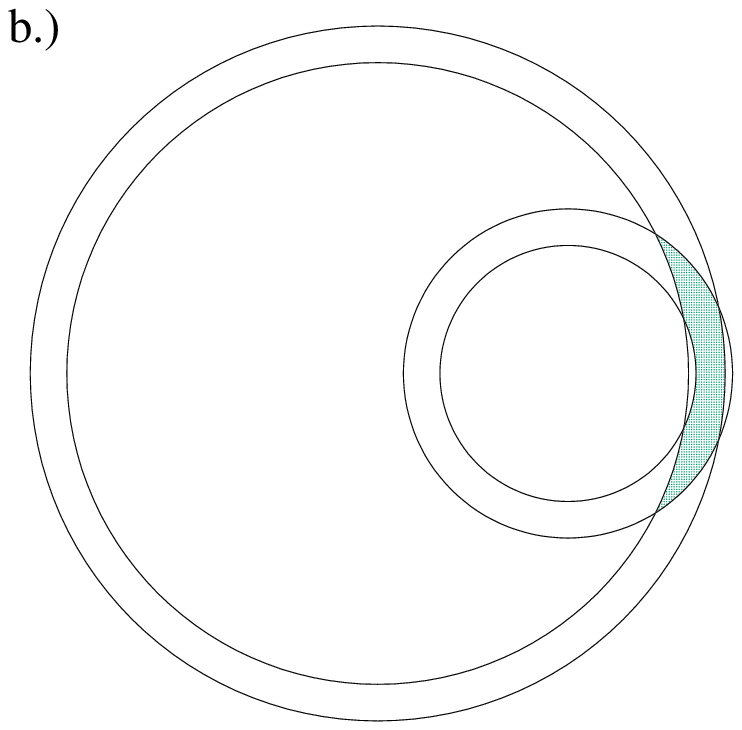}
\caption{Examples of the $r_{2}$-$r_{3}$ spherical shell overlap region in two
special cases which are  automatically accounted for by equation
\eqref{eq:sphereSphereGen} but would not work with \eqref{eq:sphereSphere}.
Panel a shows a case where $r_1$ is less than $\Delta r$. Panel b shows the
cross-section for nearly co-linear triangles.}
\label{fig:specials}
\end{figure}

From careful study of Figure~\ref{fig:overlap}, we can find that the volume of
interest is given by first finding the overlap volume of the two outer
spherical surfaces, then subtracting the overlap volume of the outer
spherical surface of one shell with the inner spherical surface of the other
and vice versa. However, this ends up removing the overlap volume of the two
inner spherical surfaces twice, so we have to add one back. Mathematically,
this can be expressed as
\begin{equation}
\label{eq:shellShell}
\begin{array}{@{}l@{\;}l}
V_{\mathrm{cs}}(r_{1},r_{2},r_{3},\Delta r) =& V(r_{1},r_{2} + \Delta{r}/2,r_{3} + \Delta{r}/2)\\
                                    & {}- V(r_{1},r_{2} + \Delta{r}/2,r_{3} - \Delta{r}/2)\\
                                    & {}- V(r_{1},r_{2} - \Delta{r}/2,r_{3} + \Delta{r}/2)\\
                                    & {}+ V(r_{1},r_{2} - \Delta{r}/2,r_{3} - \Delta{r}/2).\\
                                     \end{array}
\end{equation}
Here, care must be taken that the $r$ of equation~\eqref{eq:sphereSphereGen} 
is actually the smaller of the two spherical shell
radii. For example, when considering isosceles or equilateral triangles, it is
possible that $r_{2} + \Delta r/2$ is larger than $r_{3} - \Delta r/2$, since
$r_{2}$ may  equal $r_{3}$. Exercising this caution, along with the special
considerations of equation~\eqref{eq:sphereSphereGen} will yield the correct
volume even in special cases where the overlap volume looks quite different
than in Figure~\ref{fig:overlap} -- see e.g. Figure~\ref{fig:specials}. This
volume times $n_{R}$ is the number of $p_3$ points falling into the overlap region.

For a finite bin width, $p_2$ can be anywhere inside a spherical shell
$r_1 \pm \Delta r/2$ around $p_1$ so we will have to integrate 
equation~\eqref{eq:shellShell} with respect to $4\upi n_{R} r_1^2\mathrm{d}r_1$. Finally,
we have to account for the fact that there are on average 
$n_{R}V_{\mathrm{box}}$ points $p_1$. 
Combining all of this results in an exact expression for the expected 
number of $RRR$ counts without any
shot-noise expressed as a simple one dimensional integral,
\begin{equation}
\label{eq:RRR}
    RRR(r_{1},r_{2},r_{3}) = 4\upi s n_{R}^{3}V_{\mathrm{box}}\int_{r_{1} -
    \Delta r/2}^{r_{1} + \Delta r/2} r_{1}^{\prime 2}V_{\mathrm{cs}}(r_{1}^
    {\prime},r_{2},r_{3},\Delta r)\mathrm{d}r_{1}^{\prime}.
\end{equation}
Here $s$ is the number of unique permutations of the side lengths, with $s = 1$, $3$, and $6$
for equilateral, isosceles, and general triangles, respectively. We note 
that for most triangles this reduces to a surprisingly simple expression,
\begin{equation}
\label{eq:RRRsimple}
RRR(r_{1},r_{2},r_{3}) = 8\upi^{2} s n_{R}^{3}V_{\mathrm{box}}r_{1}r_{2}r_{3}\Delta r^{3}.
\end{equation}
However, this expression will break down in special cases such as those shown
in Figure~\ref{fig:specials}. For this reason, we recommend simply evaluating the
integral in equation \eqref{eq:RRR} numerically.\footnote{The terms in equation \eqref{eq:shellShell}
will at most be polynomials of degree 5 in $r_{1}^{\prime}$, meaning a simple 3
point Gaussian quadrature rule is all that is needed.}

\subsection{DRR counts}

For $DRR$ counts we have to replace the first vertex by a data point. Since
data and random points are uncorrelated\footnote{This would 
not be the case for nontrivial
survey geometries where the exact placement of data points with respect to the
boundary makes a difference.} between each other the formula for the
$DRR$ counts coincides with the expression for $RRR$ in equation
\eqref{eq:RRR} with one minor change due to the potentially differing number
densities of data and random points
\begin{equation}
\label{eq:DRR}
    DRR(r_{1},r_{2},r_{3}) = 4\upi s n_{R}^{2}\overline{n}_{D}V_{\mathrm{box}}\int_{r_{1} -
    \Delta r/2}^{r_{1} + \Delta r/2} r_{1}^{\prime 2}V_{\mathrm{cs}}(r_{1}^
    {\prime},r_{2},r_{3},\Delta r)\mathrm{d}r_{1}^{\prime},
\end{equation}
where $\overline{n}_{D}$ is the average number density of data points --
i.e. $\overline{n}_{D}~=~N_{D}/V_{\mathrm{box}}$.

\subsection{DDR counts}
The procedure for predicting the $DDR$ counts is almost identical to that outlined in
section \ref{subsec:RRR}, except
for the last step where we integrate over $4\upi n_{R} r_1^2dr_1$. Since the data
points are clustered, the distribution of $p_2$ around $p_1$ is not uniform and
the number density depends on the distance $r_1$ resulting in
\begin{equation}
\label{eq:DDR}
    \begin{aligned}
    DDR(r_{1},r_{2},r_{3}) =&\; 4\upi n_{R}\overline{n}_{D}V_{\mathrm{box}}\;\times \\
                            &\quad\left\lbrace \displaystyle\int_{r_
                            {1} - \Delta r/2}^{r_{1} + \Delta r/2}n_{D}
                            (r_1^\prime)r_{1}^
                            {\prime 2}V_{\mathrm{cs}}(r_{1}^{\prime},r_{2},r_{3},\Delta r)\mathrm{d}r_{1}^{\prime}\right.\\
                            &\left.\quad{}+ \vphantom{\displaystyle\int_{r_{1} - \Delta r/2}^{r_{1} + \Delta r/2}}\mathrm{permutations}\right\rbrace{.}
                            \end{aligned}
\end{equation}
\noindent
Here $n_{D}(r)$ is the nonuniform data number density, which can easily be found 
from the data pair counts by first computing the number density in each spherical
shell
\begin{equation}
n_{D,i} = \dfrac{3DD_{i}}{4\upi N_{D}[(r_i + \Delta r/2)^3 - (r_i - \Delta
r/2)^3]},
\end{equation}
and then linearly interpolating\footnote{Higher order interpolation schemes
should also be fine to use here, but due to the need to also extrapolate in
the first and last shell, linear order was used for simplicity.} to the value
at the particular $r$ needed for the numerical integration.

Note that we can no longer use a simple integer multiple to account for the
unique permutations of the side lengths as the value of $n_{D}(r)$ will change
as we permute. Given this, there will be 1, 3 or 6 terms between the braces for 
equilateral, isosceles or general triangles, respectively. We note that when summing these
terms, it is necessary to store the number of triangles as a floating point
number to ensure accuracy. After the summing of all the needed terms in the
braces, the number can be rounded to an integer if desired.

Equations~\eqref{eq:RRR},~\eqref{eq:DRR},~and~\eqref{eq:DDR} express all terms 
in the estimator of equation~\eqref{eq:estimator} that involve random points 
without any shot-noise and in terms of simple one-dimensional integrals. $DDR$ 
counts do require the pair counts of data points, but by computing these
along with the $DDD$ counts, no additional computational time is required. For 
the benefit of the community, we have publicly released our code for these predictive calculations.\footnote{\href{https://github.com/dpearson1983/rawor}{https://github.com/dpearson1983/rawor}\\
Currently, this library is available for use in C++ and Python, though we note
that it is necessary to have the boost-python library to make use of the 
Python version.}

\begin{figure*}
\includegraphics[width=1.0\linewidth]{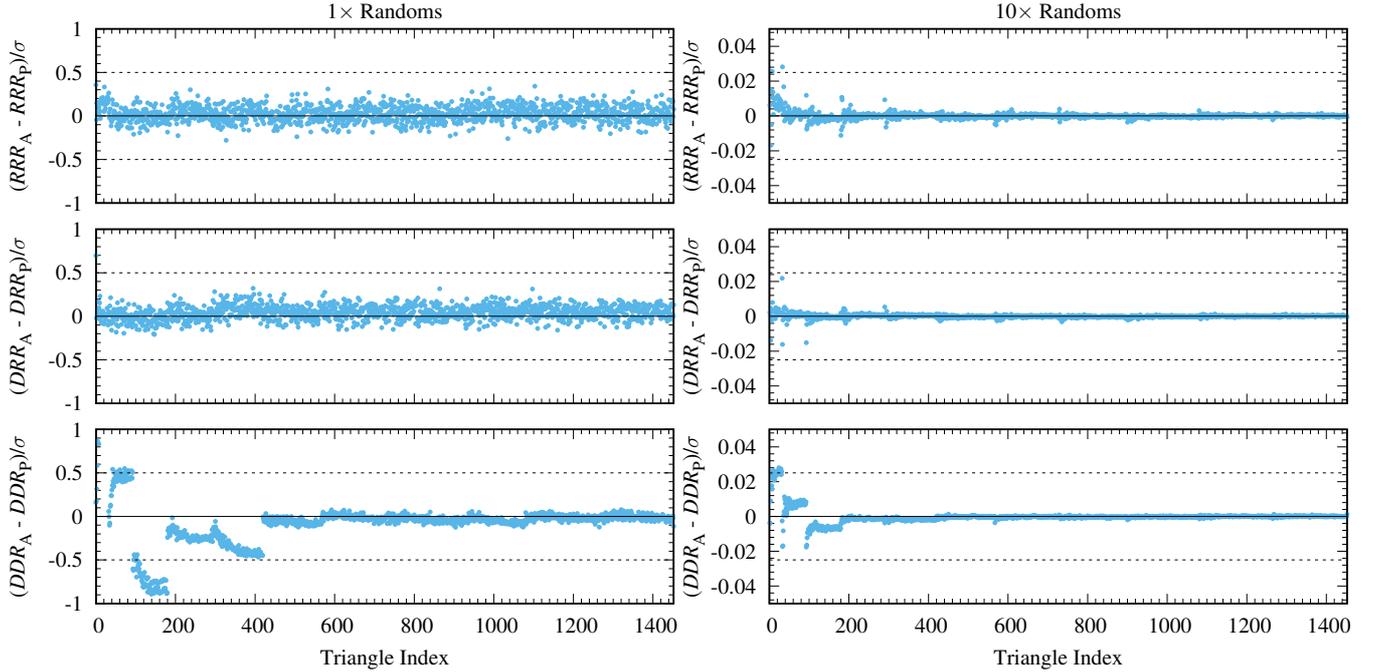}
\caption{Differences in the average actual and predicted counts divided by their 
standard deviations added in quadrature for the cases of 1$\times$ and 10$\times$ 
randoms. Note the difference in vertical scale between the left and right-hand
panels. We truncated triangle index range from 3128 to 1452. We note that the 
remainder of the range looks very similar to the range between ${\sim}600$ and 1452. 
It is clear that the $RRR$ (top) and $DRR$ (middle) predictions agree quite well 
with the predictions for both the 1$\times$ and 10$\times$ cases. The $DDR$ (bottom) 
counts agree quite well for the vast majority of triangle indices, though there are
notable deviations at lower triangle indices. We note that these effects  are not
systematic and decrease with increased number density of randoms, as well as with
increased grid resolution in the generation of the lognormal mocks.}
\label{fig:comp}
\end{figure*}

\section{Results}
\label{sec:results}
To test equations \eqref{eq:RRR}, \eqref{eq:DRR} and \eqref{eq:DDR} we compared 
their predictions with actual triplet counts. We generated 100 lognormal mock 
catalogues \citep{Coles1991} -- 
$N_{D}\sim 500000$, $V_{\mathrm{box}} = (1024~\mathrm{Mpc})^{3}$ --  using the method 
described in Appendix A of 
\cite{Beutler2011} with a power spectrum from the \textsc{camb} software 
\citep{Lewis2000} using the \emph{Planck} cosmology \citep{Planck2018} to serve as 
proxy for data points\footnote{While lognormal mocks do not adequately reproduce 
the three-point clustering statistics observed in simulations or the real Universe, 
they do contain a non-zero three-point signal which is sufficient for the purposes 
of our testing. As we already had many lognormal mocks at our disposal, we utilized 
some of them out of convenience.}, and two sets of random distributions, 100 at 
$1\times$ and 100 $10\times$ the density of the data. We 
performed the direct counts in bins of width $\Delta r = 1~\mathrm{Mpc}$, 
for $0~\mathrm{Mpc} \leq r \leq 32~\mathrm{Mpc}$, and 
calculated our predictions. Finally, we separately averaged together all 100 counts 
and predictions in order to make our comparisons.

To perform the actual counts, we used a relatively straightforward, GPU
accelerated, $\mathcal{O}(N^3)$ counting algorithm with periodic boundary
conditions to remove any edge effects along with a few simple optimizations.\footnote{
While we do not release our exact code used here publicly, we utilized the same
algorithm in a library for use in a different project. For those who may be interested,
you can view the GPU implementation at 
\href{https://github.com/dpearson1983/ganpcf/blob/master/source/ganpcf.cu}
{https://github.com/dpearson1983/ganpcf/blob/master/source/ganpcf.cu}, lines 353 -- 396 and 426 -- 561.}
We first bin the particles to a grid with spacing
equal to  $r_{\max}$, the maximum separation considered for any of the sides
of the triangle, allowing us to limit our triplet searching to nearby particles. We also
verify that the distance between
the first and second vertex is less than $r_{\max}$ before checking for the
third vertex. While it would usually be faster to not
repeat count the same triangle by simply relabeling which is the first,
second and third vertex, due to the peculiarities of GPU memory accesses, we
found that it was more efficient to do the repeat counting.

Figure \ref{fig:comp} shows the difference of the average $RRR$ (top), $DRR$ 
(middle) and $DDR$ (bottom) counts with the predictions from equations 
\eqref{eq:RRR}, \eqref{eq:DRR} and \eqref{eq:DDR} divided by their standard 
deviations added in quadrature, with the $1\times$ and $10\times$ randoms cases
on the left and right, respectively. The predictions match the actual counts remarkably 
well, especially as the number of random points increases.

\begin{figure}
\includegraphics[width=1.0\linewidth]{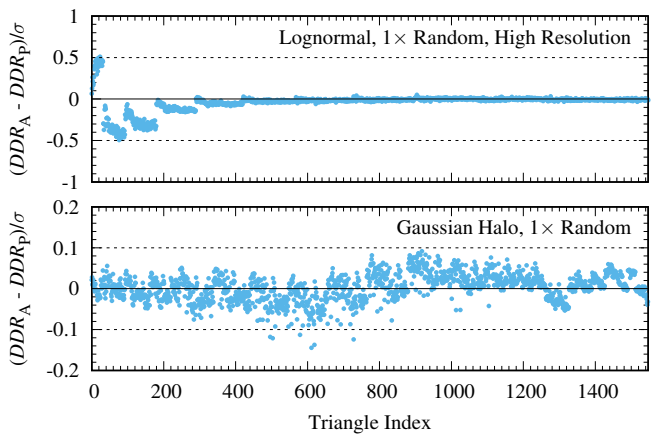}
\caption{Results of the high resolution and Gaussian halo tests for the $DDR$
predictions. The high-resolution tests (top) show that
the grid spacing is partly to blame for the deviations. The Gaussian haloes (bottom)
do not use a grid for their creation, and have an exactly known
number density profile, instead of one estimated from the data-data pair counts. We
see that there are no seemingly systematic deviations at small triangle indices. We
note that there are some slight deviations at high indices, particularly if a finer
binning is used, suggesting that the deviations have more to do with the 
discrete nature random catalogues and grid effects.}
\label{fig:DDRtests}
\end{figure}

The $DDR$ counts tend to show more significant deviations between the direct counts and our
predictions at small triangle indices -- i.e. small scales. We verified that these
deviations are not systematic -- i.e. they will fluctuate up and down with equal
probability for independent realisations averaging to zero -- and that they tend to
zero in the limit of high number density. They look systematic because at small
triangular indexes the measurements share the same small number of points and are
strongly correlated -- or anti-correlated depending on the shapes.  Additionally, the
deviations all occur when one side of the triangle is shorter than 5~Mpc, or 3~Mpc
for the $1\times$ and $10\times$ randoms cases, respectively. The lognormal mocks
were created using a grid spacing of 2~Mpc, where the number of galaxies was
determined for each grid cell, then placed inside that cell in a uniform random
fashion. We ran tests on mocks with a grid spacing of 0.5~Mpc but the same number
density, and note a reduction in the deviations -- see Figure \ref{fig:DDRtests}.

To further test that these small scale deviations were due to
some combination of low number densities and grid effects in 
the mocks, we generated haloes with an exactly known, Gaussian number density profile. 
These were purely artificial constructions where points were placed randomly a 
distance $r$ from the center with a probability proportional to the integral of the 
Gaussian function.\footnote{This gives you a function describing the number of points 
as a function of $r$ instead of just the number density.} We then calculated the 
number density in thin spherical shells to verify that it followed the input Gaussian 
function. Since we then knew the exact number density profile, this eliminated the 
need to use the data-data pair counts so long as our $p_{1}$ was taken as the central 
particle of the halo. We could then use the Gaussian that described the number density 
in the integrals equation \eqref{eq:DDR} and simply replace 
$\overline{n}_{D}V_{\mathrm{box}}$ with 1. The random catalogues were set to the same 
number density -- e.g. $N_{D} = N_{R} = 500000$. The number density profile was such 
that particles were at most ${\sim}25$~Mpc from the center with very few particles in 
the outer parts of the haloes, which led to seeing a similar, though significantly 
smaller, seemingly systematic effect for high triangle indices.

We show the results of using a higher resolution grid for creating the lognormal mocks 
and the Gaussian haloes in Figure \ref{fig:DDRtests}. These results suggest that the 
seemingly systematic deviations in the $DDR$ counts versus predictions are a combined 
result of small numbers of data-data pairs causing correlations in the counts, and mock
grid resolution, not a failing of the algorithm presented here.

\section{Conclusions}

We present a method for predicting the counts involving a random catalogue for
the 3PCF analysis of simulated or mock data that is free of shot-noise and
does not require random catalogues. We have shown that the
predictions from equations~\eqref{eq:RRR},~\eqref{eq:DRR},~and~\eqref{eq:DDR} agree remarkably
well with the actual counts, while keeping the same computational complexity
for arbitrarily high number densities of randoms.

The method only works for uniform periodic cubes which may lead the reader to
believe that the method is not useful for the analysis of real survey data.
This, however, is not true as there are many stages in the survey
data analysis that require computing the 3PCF from a large number of
periodic cubes.

The most obvious example is the validation of theoretical templates on $N$-body
simulations. To make sure that theoretical predictions of the 3PCF are sufficiently
accurate -- and to calibrate any systematic effects if they are not --
they must be compared to the measurements from
$N$-body simulations with a known cosmology, or a hidden cosmology in the case
of data challenges. For these validation tests to be meaningful, they need to
be performed on periodic cubes to cleanly separate theoretical systematics
from other observational effects. Since the cumulative volume of $N$-body
simulations used for this purpose needs to be much larger than the size of the
data, the method presented in this paper can save a significant amount of
computational time.

\section*{Acknowledgements}
We wish to thank Federico Marulli, Michele Moresco and Zachary Slepian for
their very insightful and useful comments on an earlier draft of this paper.
LS is grateful for the support from NASA grant 12-EUCLID11-0004 and the DOE
grant DE-SC0011840.  NASA's Astrophysics Data System Bibliographic Service and
the arXiv e-print service were used for this work. Additionally, we wish to
acknowledge \textsc{gnuplot}, a free, open-source plotting utility which was
used to create all of our figures.




\bibliographystyle{mnras}
\bibliography{predictRandoms} 




%
%


\bsp	
\label{lastpage}
\end{document}